\documentclass[12pt,letterpaper]{article}

\usepackage{amsmath,amssymb,calc}
\usepackage{graphicx}
\usepackage[nosort]{cite}

\newcommand\be{\begin{equation}}
\newcommand\ee{\end{equation}}
\newcommand{\bea}{\begin{eqnarray}}
\newcommand{\eea}{\end{eqnarray}}
\newcommand{\half}{\frac{1}{2}}
\newcommand{\hlf}{\frac{1}{2}}
\newcommand{\qrt}{\frac{1}{4}}
\newcommand{\C}[1]{$(\ref{#1})$}
\newcommand{\eq}[1]{(\ref{#1})}

\def\tr{{\rm tr\,}}
\def\Tr{{\rm Tr\,}}
\newcommand{\ap}{\alpha'}

\newcommand{\nn}{\nonumber}
\newcommand{\pd}{\partial}

\def\id{\protect{{1 \kern-.28em {\rm l}}}}

\def\cF{{\cal F}}
\def\cR{{\cal R}}

\def\cN{{\cal N}}

\def\cV{{\cal V}}

\def\cM{{\cal M}}
\def\M{{\cal M}}
\def\H{{\cal H}}

\def\cG{{\cal G}}
\def\L{{\cal L}}
\def\D{\Delta}
\def\G{\Gamma}

\def\Om{\Omega}
\def\la{\lambda}
\def\nab{\nabla}
\def\dd{\delta}
\def\del{\partial}
\def\a{\alpha}
\def\e{\epsilon}
\def\g{\gamma}
\def\m{\mu}
\def\n{\nu}
\def\k{\kappa}

\def\o{\omega}
\def\delbar{\bar{\partial}}
\newcommand{\ibar}{{\bar \imath}}
\newcommand{\jbar}{{\bar \jmath}}
\def\1{^{(1)}}
\def\0{^{(0)}}
\def\2{^{(2)}}
\def\f{\phi}

\def\d{{\rm d}}
\def\Re{{\rm Re}\,}
\def\Im{{\rm Im}\,}
\newcommand{\Spin}{\operatorname{Spin}}

\def\t{\tau}

\font\mybb=msbm10 at 12pt

\def\bb#1{\hbox{\mybb#1}}

\def\Z {\bb{Z}}

\newcommand{\R}{{\mathbb R}}

\def\id{\protect{{1 \kern-.28em {\rm l}}}}

\let\non\nonumber

\begin{document}
\begin{titlepage}

\begin{center}

{August 27, 2010} \hfill         \phantom{xxx} \hfill EFI-10-10

\vskip 2 cm {\Large \bf The Leading Quantum Corrections to Stringy }
\vskip 0.3 cm {\Large \bf K\"ahler Potentials}\non\\
\vskip 1.25 cm {\bf  Lilia Anguelova$^{a}$\footnote{anguella@ucmail.uc.edu}, Callum Quigley$^{b}$\footnote{cquigley@uchicago.edu} and Savdeep Sethi$^{b,c}$\footnote{sethi@uchicago.edu}}\non\\
{\vskip 0.5cm  $^{a}$ {\it Dept. of Physics, University of Cincinnati,
Cincinnati, OH 45221, USA}\non\\ \vskip 0.2 cm
$^{b}$ {\it Enrico Fermi Institute, University of Chicago, Chicago, IL 60637, USA}\non\\ \vskip 0.2cm
$^{c}$ {\it Institute for Theoretical Physics, University of Amsterdam,
Valckenierstraat 65, 1018 XE Amsterdam, The Netherlands} \non\\}

\end{center}
\vskip 2 cm

\begin{abstract}
\baselineskip=18pt

The structure of stringy quantum corrections to four-dimensional
effective  theories is particularly interesting for string
phenomenology and attempts to stabilize moduli. We consider the
heterotic string compactified on a Calabi-Yau space. For this case,
we compute the leading corrections to the kinetic terms of moduli
fields.  The structure of these corrections is largely dictated by
the underlying higher-dimensional extended supersymmetry. We find
corrections generically of order ${\alpha'}^2$ rather than of order
${\alpha'}^3$ found in type II compactifications or heterotic
compactifications with the standard embedding. We explore the
implications of these corrections for breaking no-scale structure.


\end{abstract}

\end{titlepage}

\tableofcontents

\section{Introduction}

A primary question in string theory is to understand the structure of the four-dimensional effective theories that emerge from supersymmetric compactifications. Are they generic N=1 field theories coupled to supergravity or does the requirement of consistent coupling to quantum gravity impose constraints on the field content and couplings? For models with large volume limits, the four-dimensional field content and basic interactions can be determined using supergravity. These interactions can be organized into a space-time K\"ahler potential and a superpotential. To go beyond the basic supergravity approximation requires an understanding of string quantum corrections which depend on the string scale $\alpha'$. These corrections can renormalize both the space-time K\"ahler potential and superpotential. In the past, some of these quantum corrections have been beautifully computed for world-sheet theories with $(2,2)$ supersymmetry using mirror symmetry~\cite{Candelas:1990rm}, and by studying string threshold corrections to gauge kinetic terms~\cite{Dixon:1989fj, Dixon:1990pc}.

The focus in N=1 theories is typically  on space-time superpotential
couplings which are strongly constrained by holomorphy. Much less is
understood about the K\"ahler potential which appears to be less
constrained. We suspect, however, that much more can be said about
stringy K\"ahler potentials largely through a heterotic
generalization of mirror symmetry to $(0,2)$ rather than $(2,2)$
world-sheet theories~\cite{Blumenhagen:1996vu, Blumenhagen:1996tv,
Blumenhagen:1997pp, Adams:2003zy, Katz:2004nn, Sharpe:2006qd,
Guffin:2007mp, McOrist:2008ji, Melnikov:2010sa}. One of the basic
tests of the original $(2,2)$ mirror conjecture was reproducing the
known $(\alpha')^3$ correction to the space-time prepotential of N=2
type II compactifications~\cite{Candelas:1990rm}. We would like an
analogous understanding of the leading quantum correction to stringy
K\"ahler potentials determining the kinetic terms of moduli fields.
This will provide data to help formulate a precise $(0,2)$ mirror
map.

In type II theories, the $(\alpha')^3$  correction to the moduli
kinetic terms can be understood both from a sigma model
computation~\cite{Grisaru:1986px}\ and from a space-time perspective
by reducing the $R^4$ terms in ten dimensions on the
compactification space~\cite{Antoniadis:1997eg}. These $R^4$ terms
are down by $(\alpha')^3$ from the supergravity couplings but their
form and moduli dependence is determined completely by
supersymmetry~\cite{Green:1998by}. Similarly, higher-dimensional
supersymmetry determines the leading correction to the K\"ahler
potential of N=1 compactifications. For example, in the
heterotic/type I string, there are $R^2$ corrections to the
supergravity interactions in ten dimensions suppressed by $\alpha'$.
These couplings are also completely determined by space-time
supersymmetry~\cite{Bergshoeff:1989de}. It is these couplings that
give rise to the leading quantum corrections to the kinetic terms in
four dimensions.

The setting we will consider is the weakly  coupled heterotic string
on a Calabi-Yau space with some choice of bundle preserving N=1
space-time supersymmetry. What we find is that the leading
correction to the moduli kinetic terms, in powers of the volume, is
order $(\alpha')^2$ rather than $(\alpha')^3$. It is in no way
surprising that the correction is larger than the special case of
$(2,2)$ world-sheet theories. What is perhaps more surprising is
that the correction is not $O(\alpha')$!

Now it might appear that type II  Calabi-Yau orientifolds, often
studied in moduli stabilization scenarios, are in a different class
with kinetic terms determined by projecting the underlying N=2
prepotential with its $(\alpha')^3$ perturbative correction. However
this is not the case. The orientifold planes and D-branes whose
inclusion is required for a consistent N=1 theory themselves support
$R^2$ couplings. It is these couplings that again determine the
leading quantum corrections and they will be larger than those found
in the orientifolded N=2 prepotential except if the string coupling is very small.
This clearly must be the case
since type I itself can be viewed as an orientifold of type IIB
string theory by world-sheet parity but the ten-dimensional type I
space-time action contains exactly the same $R^2$ terms as the
heterotic string. For more general orientifolds of a Calabi-Yau, the
moduli dependence of the leading correction might differ from the
heterotic/type I cases but the order of the correction in powers of $\ap$ will be the
same.

A natural application of these results  is to moduli stabilization.
In type IIB and more general F-theory compactifications to four
dimensions, many moduli of the underlying Calabi-Yau geometry are
stabilized by $G_3$ fluxes~\cite{Dasgupta:1999ss}. In principle, the
only modulus left unfixed is the overall volume. At large volume,
this lifting of moduli can be described by a space-time
superpotential~\cite{Gukov:1999ya}. In the heterotic string, there
is a dual version of this lifting which involves turning on
torsion~\cite{Dasgupta:1999ss}. However, it is a misnomer to call
this moduli lifting because heterotic compactifications with flux
are topologically distinct from those without flux. Again, in
principle, there is only one modulus unfixed which is the heterotic
dilaton.

The background we consider certainly contains flux via the heterotic
gauge bundle but it does not contain torsion except at sub-leading
orders in $\alpha'$. What we gain by this restriction is a large
volume limit where we can trust an $\alpha'$ expansion. What we lose
is a dilaton with large variation at tree-level (dual to a large
warp factor in type IIB). It would be interesting to see if
there are yet larger quantum corrections for more general fluxes
that include torsion, or perhaps for more general non-geometric heterotic backgrounds~\cite{ReidEdwards:2008rd, McOrist:2010jw}.
We suspect this might not be the case since
more general fluxes, and even non-geometries, do not appear particularly distinguished in
F-theory duals on Calabi-Yau four-folds.


Still we can examine type IIB stabilization scenarios like the LARGE volume
scenario (LVS) that use the underlying N=2 kinetic terms derived from an
orientifolded prepotential~\cite{Balasubramanian:2005zx}, and ask how they change if we include
the larger perturbative quantum correction found here. In the type IIB frame, our $(\ap)^2$ correction is suppressed
by powers of $g_s$ from the $(\alpha')^3$  perturbative correction obtained from the underlying N=2 theory
by orientifolding. This $(\alpha')^3$ correction plays a prominent role in the LVS scenario. Since the string coupling in type IIB is not
typically small at a stabilized point, the leading quantum corrections computed in an N=1 framework are needed to determine whether there are stabilized solutions even in effective field theory.

We should also
stress that there are deeper unresolved issues with instanton
computations in all scenarios that attempt to stabilize the final
modulus, but we will simply neglect those issues here. They will be
explored elsewhere.\footnote{In a nutshell, brane instanton computations in string theory
are sensitive to whether supersymmetry is broken or unbroken by background fields like fluxes.
Therefore determining the dependence of instanton corrections on field directions that break
supersymmetry and fail to satisfy the supergravity equations of motion
requires off-shell information. This makes the organization of low-energy physics into a K\"ahler potential
and a superpotential  subtle.}

This paper is organized as follows:  in section~\ref{(2,2)review}, we
summarize some background material and introduce notation. In
section~\ref{reviewaction}, we review the $\alpha'$-corrected
ten-dimensional heterotic action and describe the vacuum solution to
${O} (\alpha'^2)$. We  compute the four-dimensional space-time
K\"{a}hler potential for moduli to ${ O}(\alpha'^2)$ in
section~\ref{4dEffAc}. The space-time K\"ahler potential we find takes the form,
\be
K  = - \log (\cV ) +{\ap^2\over 2\cV}\int  \tilde{h} \wedge \ast \tilde{h}  +O(\ap^3), \label{kahlersummary}
\ee
where $\cV$ is the volume of the Calabi-Yau space and $\tilde{h}$ is a moduli-dependent two-form.\footnote{Throughout this paper, we work in units where $\k^2 = 8\pi/M_p^2 =1.$ Of course, factors of $\k$ can always be restored on dimensional grounds.}
Finally in section~\ref{Implications}, we explore
implications of this correction for breaking no-scale structure. Included in this discussion is a comparison of
our results with complimentary computations of the string one-loop correction
to the K\"{a}hler potential in a class of type IIB orientifolds~\cite{Berg:2005ja}. We find that the $\ap^2$ correction in~\C{kahlersummary}\  breaks the no-scale structure present in heterotic and type I compactifications but not in F-theory type orientifolds.


\section{Review of the $(2,2)$ K\"{a}hler Potential}\label{(2,2)review}
\setcounter{equation}{0}

Before studying more general $(0,2)$ backgrounds, it will be useful
to first  briefly review the more familiar case where the
world-sheet theory enjoys $(2,2)$ supersymmetry. The corresponding
space-time picture is given by heterotic strings on a Calabi-Yau
background $\cM$ with the spin-connection embedded in the gauge
connection. This material is well-known and can be found in
standard textbooks.


Compactification on a Calabi-Yau manifold, $\cM$,  typically results
in many massless scalar fields in four dimensions whose vacuum
expectation values remain unfixed. Let us use $x$ to denote
space-time coordinates (indices: $\mu, \nu,
\ldots$) and $y$ for internal coordinates (indices: $m,n, \ldots$). In every
compactification, there is an axio-dilaton field \be S = a +
ie^{-2\f_4} \ee where the axion $a$ is the dual to the space-time
NS-NS 2-form $B_{\m\n}$. The $4$-dimensional dilaton is related to
the $10$-dimensional dilaton via $e^{-2\f_4}=e^{-2\f}\, \cV$, where
$\cV$ is the total volume of the internal space.

In addition to $S$, there are metric moduli with a locally split
moduli space  corresponding to deformations of the K\"ahler class
and deformations of the complex structure of $\M$. By definition,
moduli are deformations of the vacuum solution that preserve the
equations of motion. Therefore for the internal directions,
$R_{mn}=0$ must be satisfied both for the starting undeformed metric
$g$ and for the deformed one $g+ \dd g$. At leading order in the
deformation, this imposes the constraint \be \D_L \dd g_{mn} = 0 \,
, \ee where $\D_L$ is the Lichnerowicz operator; geometric moduli
are therefore zero-modes of Lichnerowicz. We will discuss this
operator in more detail in Section \ref{background-solution}.

In a complex basis, the two kinds of metric perturbations are those
with mixed indices (K\"ahler deformations) and those with indices of
pure type (complex structure deformations). To the metric
deformations of pure type, we can associate a set of $(2,1)$ forms
$\chi_I$ with $I = 1,..., h^{2,1}$ by contracting with the
holomorphic 3-form $\Om$: \be \dd g_{\ibar\jbar}(x,y) =
-{\Om_\ibar^{\ k\ell}\over\parallel\Om\parallel^2}\,\chi_{I\,
\jbar\, k \ell}\,(y)\, \dd Z^I(x) \, , \label{complex structure} \ee
where $Z^I$ are complex scalars parameterizing the deformations of
the complex structure, and similarly for the conjugate terms.
Furthermore, if $\D_L \dd g_{\ibar\jbar} = 0$ then $\D_{\delbar}\,
\chi_I = 0$. So, in fact, we have a $1-1$ map between the zero-modes
$\dd g_{\ibar\jbar}$ of Lichnerowicz and the harmonic
representatives $\chi_I$ of $H^1(T\cM)\cong H^{2,1}(\cM)$.

Similarly, the mixed type  deformations \be \dd\widetilde{ g}
= i \dd g_{i\jbar}\, \d y^i\wedge\d y^\jbar \, , \ee can be
associated to harmonic representatives of $H^1(T^*\cM)\cong
H^{1,1}(\cM)$. Indeed, let $\{ \o_\a \}_{\a = 1,...,h^{1,1}}$ denote
a basis of the integral cohomology group $H^{2}(\cM,\Z)$. Then
expanding the K\"ahler form in this basis $J = t^\a \o_\a$ gives \be
g_{i\jbar}(x,y) = t^\a(x) \o_{\a\, i\jbar}(y) \, , \ee where the
$h^{1,1}$ fields  $t^\a$ are the imaginary parts of the K\"ahler
moduli chiral fields. These fields are complexified because of the
presence of the NS-NS $2$-form field $B$. We can expand the
$B$-field in the same basis $\o_\a$ for $H^{2}(\cM, \Z)$  \be
B_{i\jbar}(x,y) = b^\a(x) \o_{\a\, i\jbar}(y) \ee and define
$T^\alpha$ via \be B+iJ = T^\a\o_\a \, , \label{complexified Kahler}
\ee where the $h^{1,1}$ complex scalars, \be T^\a=b^\a+it^\a, \ee
are the K\"{a}hler moduli.

An important fact that we will  need later is the existence of
$h^{1,1}$ Peccei-Quinn symmetries, \be b^\a \mapsto b^\a + \e^\a \,
, \ee which are preserved to all orders in $\ap$ perturbation theory
regardless of the string coupling. These PQ symmetries can be broken
by world-sheet instantons but such non-perturbative effects are not
visible in the reduction of the ten-dimensional space-time action.
These shift symmetries imply that the K\"ahler potential must be a
function of only the imaginary combination $$T^\a-\bar{T}^\a$$ in
$\alpha'$ perturbation theory.

The classical K\"ahler potential for all of  the above moduli has
the following form: \be K_{cl} = -\log\left(-i(S-\bar{S})\right)
-\log\left( i\int \Om\wedge\bar{\Om}\right) - \log\left(\int J\wedge
J \wedge J\right).\label{(2,2)potential} \ee Notably, this potential
factorizes between the three sectors: dilaton, complex structure,
and K\"ahler class. When perturbative worldsheet effects are
included, as shown in~\cite{Candelas:1990rm}, there is a
unique correction to the K\"ahler moduli sector. Namely \be \cV =
{1\over6}\int J\wedge J\wedge J \,\, \longmapsto \,\, \cV +
\ap^3\zeta(3)\chi(\cM) \label{(2,2)correction} \, , \ee where
$\chi(\cM)$ is the Euler number of $\cM$. This leads to the
following modification of the K\"{a}hler potential: $$- \log
\left(\int J\wedge J \wedge J\right) \,\rightarrow\, - \log {\cal V}
+ \alpha'^3 \, \frac{ \rm const}{{\cal V}}.$$ We would like to
understand the analogue of this result when the world-sheet theory
has only $(0,2)$ SCFT. From a space-time perspective, this means
that we are interested in heterotic compactifications with
non-standard embedding and space-time GUT groups other than just
$E_6$.

\section{The Quantum-Corrected Background} \label{reviewaction}
\setcounter{equation}{0}
\subsection{Ten-dimensional space-time action}
\label{action}

The heterotic effective action is fixed to ${O}(\alpha'^2)$ by supersymmetry. However, there are different conventions for the choice of connections, used to evaluate curvatures. A particularly natural one leads to the following space-time effective action~\cite{Bergshoeff:1989de, Bergshoeff:1988nn}:
\begin{equation} \label{HetAc}
S=\frac{1}{2\kappa_{10}^2} \int d^{10}x \sqrt{-g} \, e^{-2 \Phi} \Big[\cR
+4 (\partial \Phi)^2-\frac{1}{2} |{\cal H} |^2  -
\frac{\alpha'}{4} \left( \tr |{\cal F}|^2 -\tr |
\cR_+|^2\right)   +O(\alpha'^3) \Big],
\end{equation}
where
\begin{equation}
\tr | \cR_+|^2={1\over 2} \cR_{MNAB} (\Omega_+) \cR^{MNAB}
(\Omega_+)
\end{equation}
with $M$,$N, \ldots$ running over all ten dimensions. Also, $ {\cal F}$ is the Yang-Mills field strength and $\Phi$ denotes the $10$-dimensional dilaton.

The Einstein-Hilbert term is constructed using the Levi-Civita
connection, while the Riemann tensor appearing in the
${O} (\alpha')$ correction is built using the connection $\Omega_+$,
where
\begin{equation}\label{conn}
{\Omega_\pm}_M= {\Omega}_M\pm \half {{\cal H}}_M
\end{equation}
and ${\Omega}$ is the spin connection. The definition of ${\cal H}$ already
includes ${O} (\alpha')$ corrections,
\begin{equation}\label{aaav}
{\cal H} = d {\cal B} +\frac{\alpha'}{4}   \left[ {\rm CS}(\Omega_+) - {\rm
CS}(A)  \right] ,
\end{equation}
where $A$ is the connection on the gauge-bundle and $CS$ denotes the Chern-Simons invariant. The associated Bianchi identity is given by \be \label{bianchi}
d {\cal H} =  { \alpha' \over 4}  \Big\{
 \tr [\cR(\Omega_+) \wedge \cR(\Omega_+)] - \tr [\cF \wedge \cF] \Big\} ,
\ee
where ${\cal H}$ satisfies the quantization condition
\be
 {1\over 2\pi \alpha'} \int {\cal H} \in 2 \pi \Z \,\, .
\ee
This choice of connections is very convenient for comparison with results, obtained by T-duality from type IIB backgrounds~\cite{Becker:2009df, Becker:2009zx}. But more importantly, with this choice of fields there are no purely bosonic couplings at ${O} (\alpha'^2)$, other than those in $|{\cal H}|^2$, as we have indicated in~\C{HetAc}.

At ${O} (\alpha'^3)$ there are $R^4$ type terms whose form is not determined by supersymmetry.  However, as we will see later, the ${O} (\alpha'^2)$ terms already give the first non-vanishing correction to the K\"{a}hler potential of the four-dimensional effective theory. So, for our purposes, we will not need to look at higher orders. Presumably an answer from $(0,2)$ mirror symmetry will provide information to all orders in $\alpha'$.

The equations of motion resulting
from the action with bosonic terms~\C{HetAc}\ take the form:
\bea \label{eom}
 \cR-4 (\nabla \Phi)^2+4 \nabla^2 \Phi -{1\over 2 } \mid {\cal
H}\mid^2-{\alpha'\over 4} \left( \tr \!\!\mid {\cal F} \mid^2 - \,\tr \!\!\mid
\cR_+\mid^2\right) &=& {O} (\alpha'^3)\, , \cr
 \cR_{MN}+2 \nabla_M \nabla_N
\Phi -{1\over 4} {\cal H}_{MAB} {{\cal H}_N}^{AB} ~~~~~~~~~~~~~~~~~~~~~~~~~~~~~~~~&&\\
-{\alpha'\over 4} \Big[\tr \cF_{MP}{\cF_N}^P   -\cR_{MPAB}(\Omega_+)
\cR_{N}^{~~PAB}(\Omega_+)\Big] &=& {O} (\alpha'^3)\, , \cr
 \nabla^M \left( e^{-2
\Phi} {\cal H}_{MNP}\right) &=& {O} (\alpha'^3)\, , \cr
 D^{(-)M} (e^{-2 \Phi} {\cF}_{MN})& =& {O} (\alpha'^3)\, , \non
\eea
where $D^{(-)}=\nabla^{(-)} + [A,\,\cdot\,]$ is the gauge-covariant derivative with respect to both the gauge and the $\Om_-$ connections.


In addition to satisfying the equations of motion, we demand that our solutions preserve supersymmetry.
The supersymmetry variations of the fermionic fields appearing in ten-dimensional ${ N}=1$ supergravity lead to the Killing spinor  conditions for unbroken supersymmetry. The fermions consist of
the gravitino, $\Psi_M$, the dilatino, $\la$, and the gaugino, $\chi$. These are all Majorana-Weyl  fermions. In the convenient field choice of~\cite{Bergshoeff:1989de}, these variations take the form
\bea
&& \dd\Psi_M =\left(\del_M + \qrt\G_{AB}\left(\Om_{-\,M}{}^{AB} + \ap P_{M}{}^{AB}\right) +O(\ap^3)\right)\e =0 \, , \non\\
&& \dd\la = -{1\over2\sqrt{2}}\left( \pd\!\!\!/ \Phi - \frac{1}{2} {\cal H}\!\!\!\!/ + {3\over2}\alpha' P\!\!\!\!/  + {O} (\alpha'^3) \right)\epsilon =0 \, ,\label{susy}\\
&& \dd\chi = -\frac{1}{2} \cF\!\!\!/ \e + {O} (\alpha'^3)= 0\non \, ,
\eea
where
\be
P_{MAB} = 6e^{2\Phi}\nab^{(-)N}\big(e^{-2\Phi}\d\H\big)_{MNAB}
\ee
and $\H\!\!\!\!/={1\over6}\G^{MNP}\H_{MNP}$ with a similar expression for $P\!\!\!\!/$. Note that the $\alpha' P$ terms in~\C{susy}\ are of ${O} (\alpha'^2)$ because $d {\cal H}$ is ${O}(\alpha')$.

\subsection{Perturbative solution}\label{background-solution}

The $O(\alpha')$ corrections to supergravity described in section~\ref{action}\ lead to corrections of any vacuum solution of supergravity.  The corrected
solutions were first studied in~\cite{Witten:1986kg}\ and more recently in~\cite{Gillard:2003jh}.
Our goal will be to reduce the ten-dimensional heterotic action on the $\alpha'$-corrected
solutions of the equations of motion and extract the resulting
K\"{a}hler potential. We will follow more closely the approach of~\cite{Gillard:2003jh}\ which was carried out in string frame since this greatly simplifies many aspects of our calculations. At the end, we will give the four-dimensional effective action in Einstein frame.

We begin with some basic restrictions on the class of solutions we consider. We take our space-time to be a direct product
$\R^{1,3}\times\cM$, where $\cM$ is a compact manifold. For simplicity,
we assume that $\cM$ has no isometries, which is generically the case, although our analysis can be easily generalized to allow for isometries. Including gauge fields requires the choice of a holomorphic vector bundle $E\rightarrow\cM$. We also assume
that the NS flux $\H$ vanishes to leading order and is only induced as a correction at ${O} (\ap)$. This implies that at zeroth order we have a CY manifold. The non-K\"{a}hlerity caused by a non-zero $\H$ is determined by the fundamental form $J$ associated to the metric~\cite{Strominger:1986uh},
\be
\H = i(\del - \delbar)J \label{nonKahler}.
\ee
This arises only at ${O} (\alpha')$. This assumption simplifies the analysis but, more importantly, it guarantees that the supergravity approximation is reliable.


Let us now write down the ${O} (\ap)$-corrected solutions given in~\cite{Gillard:2003jh}.
In a complex basis, the solutions take the form:
\bea
{G}_{i\jbar} &=& {g}_{i\jbar} + \alpha' h_{i\jbar} \, , \label{metriccorrection}\\
\Phi &=& \f_{0} - \ap\xi h^{i}_{i} \, , \label{dilatoncorrection}\\
\H_{ij\bar{k}} &=&  {\ap}\big(-\nab_i h_{j\bar{k}} + \nab_j h_{i\bar{k}}\big) \, , \label{Hcorrection}\\
A_i &=& A\0_i+\ap A\1_i \, , \\
\cF_{i\jbar} &=& F_{i\jbar} +\ap F\1_{i\jbar} \, , \label{Fcorrection}
\eea
where $\xi$ is a gauge parameter that we will discuss momentarily. The covariant derivatives, $\nab$, here and subsequently are computed with respect to the zeroth order metric $g$. This is also the metric used to raise and lower indices. The correction to the gauge-field curvature, $F\1$, is determined by the conditions
\bea
F\1 &=& D A\1 \equiv (d+A\0\wedge)A\1 \, , \\
g^{i\jbar}F\1_{i\jbar} &=& h^{i\jbar}F_{i\jbar} \, .
\eea

At zeroth order, the background fields are the familiar ones for Calabi-Yau compactifications: namely, a Ricci-flat K\"{a}hler metric $g$, a constant
dilaton $\f_0$, vanishing $\H$, and a connection $A\0$ on $E\rightarrow\cM$ whose curvature satisfies the Hermitian-Yang-Mills equations
\be F_{ij}=F_{\ibar\jbar}=g^{i\jbar}F_{i\jbar}=0.\ee
This all follows directly from the Killing spinor equations~\C{susy}.

At ${O} (\ap)$, all correction terms of interest for us, namely for $G_{i \jbar}, \Phi$ and ${\cal H}$, can be written in terms of the metric correction $h$. This is also a direct consequence of the
supersymmetry equations~\C{susy}. For example, the $\H$-flux condition~\C{Hcorrection}, or equivalently~\C{nonKahler}, reflects the fact that
the holonomy group of $\nab^{(-)}$ is $SU(3)$, as required by the gravitino supersymmetry variation. However, supersymmetry alone is not sufficient to determine
$h$. We must study the $O(\ap)$ corrected equation of motion for $h$ following from~\C{eom}. In terms of a real basis, these equations
read
\be \label{heom}
\D_L h_{mn} + \xi\nabla_m\nabla_n h = \qrt\left[ \tr(F_{mp}F_n^{ p}) - R_{mpqr}R_{n}^{\ pqr} \right] ,
\ee
where $R_{mnpq}$ denotes the zeroth order Riemann tensor obtained from the metric $g_{mn}$, and $\D_L$ is the Lichnerowicz operator
\be \label{Lichnerowicz}
\D_L h_{mn} = -\hlf \nabla^2 h_{mn} - R_{mpnq}h^{pq} + \nab_{(m}\nab^p h_{n)p} + R_{p(m}h^{p}_{n)} - \hlf\nab_m\nab_n h
\ee
with
\be
h \equiv h^{m}_m = 2 h^{i}_i \label{htrace} \, .
\ee
Clearly, we can set $R_{mn}=0$ in~\C{Lichnerowicz}\ but we have included it for completeness.

The equations for $h_{mn}$ can be simplified if we impose the gauge fixing condition
\be
\nab^n h_{mn} = \big(\hlf-\xi\big) \nab_m h \, .\label{gauge}
\ee
A standard argument
is easily generalized to show that such a gauge choice is always permissible provided $\xi\geq0$.\footnote{The basic idea of the argument is to suppose that $h_{mn}$ does not satisfy the gauge condition~\C{gauge}, but that
there exists $h'_{mn} = h_{mn} +\nab_m v_n +\nab_n v_m$ which does. The task then is to show that a suitable $v$ always exists. This amounts to showing
that $v$ does not lie in the kernel of a certain second-order differential operator.  So long as $\xi\geq0$, it is easy to show that the kernel in question
is trivial, so $v$ is well defined. For more details, see~\cite{Witten:1986kg}}
On a Ricci-flat manifold,~\C{heom}\ then becomes
\be
-\left(\nab^2\dd^p_m\dd^q_n + 2 R^{p\,\,\,q}_{\,m\,\,\,\,n}\right)h_{pq}= \hlf\left[ \tr(F_{mp}F_n^{\ p}) - R_{mpqr}R_{n}^{\ pqr} \right] \label{heom2}
\ee
independent of $\xi$!

Often, the operator appearing on the left hand side is referred to as \textit{the} Lichnerowicz operator, even though this is only true in
the gauge $\xi=0$ and on a Ricci flat manifold. In general, the Lichnerowicz operator will contain the gauge-dependent piece
$-2\xi g^{pq}\nab_m\nab_n $. For reasons that will become clear
later, we will work almost exclusively in the $\xi=0$ gauge. For any exceptions, we will write $\xi$ explicitly.

The significance of~\C{heom}, or equivalently~\C{heom2}, is the following. These equations tell us that, to order $\ap$, we can completely fix the corrections to
the metric, and hence all the supergravity fields, as long as the Lichnerowicz operator is invertible. Said differently, if $h_{mn}$
does not contain any zero-modes of Lichnerowicz, then it is uniqely specified by
\be
h_{mn} = \hlf\D_L^{-1} \left[ \tr(F_{mp}F_n^{\ p}) - R_{mpqr}R_{n}^{\ pqr} \right].
\ee
What if $h_{mn}$ \textit{does} contain zero-modes? As we reviewed in section~\ref{(2,2)review}, the zero-modes of Lichnerowicz correspond to a finite set of
deformations. So any zero-modes contained in $h_{mn}$ can always be absorbed into a redefinition of the moduli fields. We will come back to this point later when we discuss the
moduli-dependence of the solutions.

It is worth pointing out a special case of \C{heom2} which comes from taking the trace; namely
\be \label{nabh1}
- \nab^2h = \tr|F|^2- \tr|R_+|^2 .
\ee
The integrability condition for this equation  is
\be
\int_\cM \d^6 y \sqrt{g}\  \left( \tr|F|^2- \tr|R_+|^2 \right) = 0
\ee
which is always satisfied  if the Bianchi identity~\C{bianchi} is satisfied.

Now if the source
$(\tr|F|^2- \tr|R_+|^2)$ is vanishing, as is the case for the standard embedding, the solution of $\nabla^2 h = 0$ can always be absorbed in a
redefinition of the zeroth order metric, as we pointed out in the previous paragraph.
This is just the statement that for the standard embedding there are no  ${O} (\alpha')$ corrections to the vacuum solution.

Clearly, the non-harmonic component of the metric deformation is essential for the physics of the non-standard embedding. In view of this, we will
impose the requirement that $h$ be non-zero only for a non-vanishing source on the right hand side of~\C{nabh1}. Equivalently, we require that $h$ be orthogonal
to the zero modes of the Laplacian. Since the only harmonic functions on a compact space are constant, this translates to the condition:
\be \label{h0}
\int_\cM \d^6 y \sqrt{g}\  h =0 \, .
\ee
For future use, let us also note that (\ref{h0}) can be rewritten as:
\be \label{htorth}
0 = \int_\cM \d^6 y \sqrt{g}\ 2 J^{i\jbar} h_{i \jbar} = \int_\cM *J \wedge \tilde{h} = \frac{1}{2} \int_\cM J \wedge J \wedge \tilde{h} \, ,
\ee
where we have used the definition $h=h^m_m$ and $\tilde{h}$ denotes the two-form constructed from $h$, i.e. $\tilde{h} = i \,h_{i \jbar} \, dy^i \wedge dy^{\jbar}$.

At $O(\alpha'^2)$, things are rather similar to the ${O} (\alpha')$ discussion above as far as the equations of motion are concerned. Most
relevant for us, the ${O} (\alpha'^2)$ correction to the metric $h^{(2)}$ defined by
\be \label{Galpha}
G_{i \jbar} = g_{i \jbar} + \alpha' h_{i \jbar} + \alpha'^2 h^{(2)}_{i \jbar}
\ee
is also orthogonal to harmonic forms. However, at this order there are new terms in the supersymmetry variations. In particular, the gravitino equation
requires the combination $\Om_- + \ap P$ to have $SU(3)$ holonomy, instead of the $\Om_-$ connection alone. This imposes the following relation:
\be \label{susypsi}
\big(\H -2\ap P\big)_{ij\bar{k}} = \del_j G_{i\bar{k}}-\del_i G_{j\bar{k}}\,.
\ee
In principle, at this order there could also be a non-trivial correction to the dilaton
\be
\Phi=\phi_0 +\ap \phi\1 +\ap^2 \phi\2 \, .
\ee
However,
we will now show that $\phi\2$ is a gauge artifact, much like $\phi\1=-\xi h_i^i$. To see this, let us consider the dilatino equation $\delta \lambda = 0$, which can be written as
\be \label{susylambda}
\del_i\Phi = \hlf\big(\H-3\ap P\big)_{ij\bar{k}}G^{j\bar{k}} +O(\ap^3).
\ee
Let us generalize the gauge condition~\C{gauge}\ to the following:
\be \label{Gengauge}
G^{j\bar{k}}\nab_{j}G_{i\bar{k}} = \left(1-2\xi\right)\nab_i \log|G| + \zeta \ap P_{ij\bar{k}}G^{j\bar{k}} \, .
\ee
This is a well-defined gauge choice $\forall\zeta$ provided $\xi\geq0$ and, furthermore, it includes (\ref{gauge}) at lowest order in $\alpha'$.\footnote{
At $O(\ap^2)$, this generalized gauge condition reads
\be
\nab^{\bar{k}}h\2_{i\bar{k}} - (1-2\xi)\nab_i\big(g^{j\bar{k}}h\2_{j\bar{k}}\big) =
h^{j\bar{k}}\nab_i h_{j\bar{k}} - (1/2-\xi)\nab_i\big(h_{j\bar{k}}h^{j\bar{k}}\big) + 6\zeta\nab^{\bar{\ell}}\big(\del\delbar h\big)_{i\bar{\ell}j\bar{k}}g^{j\bar{k}} \non
\ee
and it can always be chosen by the same arguments used earlier.} Now, using~\C{susypsi}\ and~\C{Gengauge}, we find that~\C{susylambda}\ becomes:
\be
\del_i\Phi = -2\xi\nab_i \log|g^{-1}G| + (\zeta-1)\ap P_{ij\bar{k}}G^{j\bar{k}} + O(\ap^3)\,.
\ee
So, by choosing the gauge $\xi=0$ and $\zeta=1$, we are left with a constant dilaton:
\be
\Phi = \f_0 +  O(\ap^3) \, .
\ee
Finally, let us note that the gauge condition~\C{Gengauge}\ is very similar to the family of gauge conditions studied in~\cite{Gibbons:1978ji}.

\section{The Four-dimensional K\"{a}hler Potential} \label{4dEffAc}
\setcounter{equation}{0}

In the previous section, we wrote down the vacuum solutions for $d=10$, $N=1$ supergravity compactified to four dimensions together with their leading $\ap$
corrections. Now we will study the Kaluza-Klein reduction of the ten-dimensional action~\eq{HetAc}\ on this background. As before, we use $x^\mu$ with $\mu=0,...,3$
for the space-time coordinates and $y^m$ for a real coordinate basis on the internal manifold $\M$. As before $y^i$, $y^{\jbar}$ will denote internal coordinates in a complex basis.

\subsection{A reduction ansatz}
\label{step1}

We begin by decomposing the ten-dimensional bosonic $N=1$ supergravity
fields, $g,B,\Phi$ into four-dimensional and six-dimensional components. The internal components do not depend explicitly on $x^\mu$, only indirectly via their dependence on the
moduli fields. The moduli we will consider are the dilaton, the $h^{1,1}$ complexified K\"{a}hler deformations $$T^\a = b^\a+it^\a$$
and the $h^{2,1}$ complex structure deformations $Z^I$. We will denote the set of moduli fields $\{ T^\a , Z^I\}$ collectively by $M^{\cal I}(x)$. With this notation,
the decomposition is given by:
\bea
ds^2 &=& \hat{g}_{\m\n}(x)dx^\m dx^\n + G_{mn}\big(y,M(x)\big)dy^m dy^n \, , \nn \\
B &=& B_{\m\n}(x)dx^\m dx^\n + B_{mn}\big(y,M(x)\big)dy^m dy^n \, , \nn \\
\Phi &=& \varphi(x) + \f\big(y,M(x)\big) \, , \label{dilatonsplit}
\eea
where $\hat{g}$ is the usual dynamical 4-d metric, $B_{\m\n}$ is the 2-form dual to the universal axion and, finally, $\varphi$ is the 4-d fluctuation of the 10-d dilaton $\Phi$.

In the previous sections, we discussed the intrinsic properties of
the fields $G_{mn}, B_{mn},$ and $\f$, with moduli independent of $x^{\mu}$. Now we are allowing the moduli to fluctuate in
space-time in order to obtain the four-dimensional effective action for these light fields. At the end of the day, we will only be interested in perturbative corrections up to order $\alpha'^2$. Regardless, it will be beneficial and more illuminating to work with the full expressions~(\ref{dilatonsplit}). We will expand in powers of $\alpha'$ at a later stage.

Although we will not study the Yang-Mills sector in the subsequent sections, let us for completeness briefly comment on its reduction as well. In ten dimensions, the gauge group $\cG$ of the heterotic theory is fixed by anomaly cancelation to
be either $\cG=(E_8\times E_8)\rtimes \mathbb Z_2$ or $\Spin(32)/\Z_2$. Upon compactification, the background requires a holomorphic vector bundle $E\rightarrow\cM$ to satisfy the Bianchi identity.
Let us denote the structure group of $E$ by $H$. If $G\times H$ is a maximal subgroup of $\cG$, then the holomorphic bundle breaks the space-time gauge group down to $G$.
The adjoint representation of $\cG$ will decompose into a sum of irreducible representations of $G\times H$, namely
\be
Adj(\cG) = (Adj(G),\textbf{1})\oplus_r R_r \oplus (\textbf{1},Adj(H)) \non
\ee
for some set of representations $R_r$.\footnote{More concretely, if we take $\cG=E_8\times E_8$ and embed $H$ into one $E_8$ factor (and for simplicity
ignore the remaining $E_8$ factor) some common cases are:
\bea
\begin{array}{ccc}
~~\underline{~~G~~}~~ & ~~\underline{~~H~~}~~ & \underline{~~~~~~~~~~~~~~~~~~~~\oplus_i R_i~~~~~~~~~~~~~~~~~~~~} \\
E_6                & SU(3)            & (\textbf{27},\textbf{3})\oplus (\overline{\textbf{27}},\overline{\textbf{3}}) \\
SO(10)             & SU(4)            & (\textbf{16},\textbf{4})\oplus (\overline{\textbf{16}},\overline{\textbf{4}}) \oplus (\textbf{10},\textbf{6})           \\
SU(5)              & SU(5)            & \hspace*{0.5cm}(\textbf{10},\textbf{5})\oplus (\overline{\textbf{10}},\overline{\textbf{5}})\oplus(\textbf{5},\overline{\textbf{10}})\oplus (\overline{\textbf{5}},\textbf{10}) .
\end{array} \non
\eea}
Therefore, the 10d gauge field will take the form
\bea
A &=& A_\m(x)  dx^\m + C_m(x)dy^m + A_m\big(y,a(x)\big)dy^m ,
\eea
where $A_\m$ is the $G$-valued four-dimensional gauge field, $C_m$ are charged matter fields transforming in the representations $R_r$, and
$A_m$ is a background connection on the bundle $E\rightarrow\cM$. This last field depends on the bundle moduli $a^{\Sigma}$, which appear as gauge singlets in spacetime.
More precisely, the moduli dependence of the background connection is given by:
\be
A_{\ibar}(x,y) = a^{\Sigma}(x)\a_{\Sigma\,\ibar}(y) \, ,
\ee
where $\a_{\Sigma}$ form a basis of $H^1(End~E)$. The bundle moduli are poorly understood currently even in the semi-classical large
volume limit where classical geometry is applicable. One of the key issues in understanding heterotic string better is an improved understanding of bundle and charged moduli.

\subsection{The quantum-corrected effective action}

Before turning to the reduction of the ten-dimensional action on the $\alpha'$-corrected background of section~\ref{background-solution}, let us make several useful observations which will simplify the subsequent computations.

First the classical moduli space factorizes into K\"ahler, complex and bundle deformations. There is a ground ring structure present at the level of the world-sheet $(0,2)$ theory which suggests that a K\"ahler/complex split might persist even including quantum corrections, despite only N=1 space-time supersymmetry~\cite{Adams:2003zy, Adams:2005tc}. The bundle moduli also appear to be split with a similar K\"ahler/complex categorization though we will ignore them in this discussion. Since we are concerned with perturbative corrections, it is reasonable to suspect that the complex structure moduli are insensitive to the leading volume corrections. So we will focus on the K\"ahler moduli.


Because of the shift symmetries of the $b^\a$ moduli reviewed in section~\ref{(2,2)review}, the K\"{a}hler potential is only a function of the combination $T^\a -{\bar T}^\a$. There is therefore no need to explicitly track the $b^\a$ moduli through the calculation; once we obtain the kinetic terms for the $t^\a$ fields, it is trivial to obtain the K\"{a}hler potential for $T^\a$ by rewriting those kinetic terms as a function of $T^\a - {\bar T}^\a$ instead of $t^\a$. We will therefore hold the moduli fields $b^\a$ constant in space-time.

A word on the gauge sector: we will not include gauge bundle moduli beyond incorporating a fixed background gauge bundle. Said differently, we fix the bundle moduli
$a^{\Sigma}$ to be constants and set the matter fields $C$ to zero. Certainly, understanding the bundle moduli sector is an important and interesting question.
However, the mixing of bundle and K\"ahler moduli via the Chern-Simons couplings in $\H$ complicates the analysis. So we will leave this generalization for a future investigation.

The technical details of the reduction of the ten-dimensional action~\eq{HetAc}\ to four dimensions are given in Appendix~\ref{reduction}. The result, in Einstein frame,
is the following four-dimensional effective action:
\bea
S_{eff} &=& {1\over2}\int\d^4x\sqrt{-g_E}\left[R_E+2{\del_\m S \del^\m\bar{S} \over (S-\bar{S})^2} -\cG_{{\cal IJ}}(M) \,\del_\m M^{\cal I} \del^\m M^{\cal J} \right]\label{eff-action}\\
&& -{\ap\over8}\int \Tr\Big(\Im\!(S)F\wedge*F -\Re\!(S)F\wedge F\Big) +O(\ap^3) \,\, , \non
\eea
where the subscript $_E$ denotes Einstein frame. The axio-dilaton $S=a+ie^{-2\f_4}$ contains the 4-d dilaton
\be \label{4dddef}
\f_4 = \varphi -\log\big[\cV\,'/\cV_0\big] +O(\ap^3)
\ee
with $\cV\,'(x) = \int\d^6y\sqrt{G}$ is the $\ap$-corrected volume and $\cV_0$ some reference volume. For future use, by $\cV$ we will denote
the volume of the underlying CY space, i.e. $\cV=\int\d^6y\sqrt{g}$.

Note that to this order in $\ap$,
the definition~(\ref{4dddef})\ yields the same (flat) cylindrical metric for the dilaton that appears classically. More importantly, the dilaton does not mix with the other
moduli. The metric on the rest of moduli space is

\be
\cG_{{\cal IJ}} \, = \, {1\over 4\cV\,'} \int d^6y \sqrt{G} \,\,\, \dd_{\cal I} G_{mp}\,\, \dd_{\cal J} G_{nq}\, \Big(G^{mn}G^{pq}-{\ap\over2}\cR^{mnpq}\Big) . \label{metric3}
\ee
Note that the functional derivatives are with respect to the K\"{a}hler and complex structure moduli, i.e. $\dd_{\cal I}={\dd/\dd M^{\cal I}}$, and that $\cR^{mnpq}$ denotes the Riemann tensor
computed with the full metric $G$; the curvature associated with $g$ is always denoted by $R^{mnpq}$.
All deviations of the effective action from the $\ap\rightarrow0$ limit reside in this metric for the moduli kinetic terms.

One can read off the order $\alpha'$ and order $\alpha'^2$ terms in (\ref{metric3}) by expanding
\be \label{GmnExp}
G_{mn} = g_{mn} + \alpha' h_{mn} + \alpha'^2 h^{(2)}_{mn} + {O} (\alpha'^3) \,\, .
\ee
For example, to $O(\ap)$ the moduli space metric is given by:
\bea \label{metric-order-ap}
\cG_{{\cal IJ}} \!\!&=& \!{1\over 4\cV} \int d^6y \sqrt{g}\Big(\dd_{\cal I} g_{mp} \dd_{\cal J} g_{nq} g^{mn}g^{pq} + \ap \big(\dd_{\cal I} h_{mp} \dd_{\cal J} g_{nq} + \dd_{\cal I} g_{mp} \dd_{\cal J} h_{nq} \big)g^{mn}g^{pq} \\
\!\!&& \!\qquad \qquad +{\ap\over2} \dd_{\cal I} g_{mp} \dd_{\cal J} g_{nq} \big(h g^{mn}g^{pq} - 2h^{mn}g^{pq} - 2g^{mn}h^{pq} -R^{mnpq}\big) \Big)\!+O(\ap^2) \, . \non
\eea
Obviously, the first term is the standard classical moduli space metric.
The $\alpha'$-correction terms provide a natural generalization, as we will discuss in the next subsection.

Before we begin considering the kinetic terms of the K\"{a}hler moduli in more detail, let us revisit the splitting of the K\"{a}hler and complex structure
moduli spaces in more detail. Unlike the case of $(2,2)$ compactifications, in $(0,2)$ there are currently no general arguments that those two moduli spaces should decouple. There is no space-time argument since
both $T^\a$ and $Z^I$ sit in $\cN=1$ chiral multiplets. At zeroth order, they do decouple and this can be seen simply from the index structure of the first term in~(\ref{metric-order-ap}). Indeed since $g$ is Hermitian, the only non-vanishing components of the inverse metric are $g^{i \jbar}$, and the moduli space metric
splits into K\"ahler $\cG_{\alpha \bar{\beta}} \sim \delta_\a g_{i \jbar} \, \delta_{\bar{\beta}} g_{k \bar{\ell}} \,g^{i \bar{\ell}} g^{k \jbar}$ and
complex structure $\cG_{\bar{I} J} \sim \delta_{\bar{I}} g_{i j} \,\delta_Jg_{\bar{k} \bar{\ell}} \, g^{i \bar{k}} g^{j \bar{\ell}}$ sectors, with no
mixing between them, $\cG_{\a\bar{I}}=0$. However, at higher orders in $\alpha'$ one should expect non-vanishing off-diagonal components
\be
\cG_{\a \bar{I}} = {1\over 2\cV\,'} \int d^6y \sqrt{G} \,\dd_{\bar{I}} G_{ik}\, \dd_\a G_{j\bar{\ell}}\, \Big(G^{i j}G^{k\bar{\ell}}-{\ap\over2}\cR^{ijk\bar{\ell}}\Big) \, , \label{mixed metric}
\ee
coupling the two sectors together.

While such mixing is an {\it a priori} possibility, it does not occur in our case
at least through
$O(\ap^2)$ for the following reasons. Recall we are only considering $\ap$ corrections to the K\"ahler moduli kinetic terms not the complex structure kinetic terms. Therefore, the full metric remains Hermitian ($G^{ij}=0$) and so any mixing must come entirely from
the curvature term $\cR_{ijk \bar{\ell}}$.\footnote{For non-K\"{a}hler complex manifolds, these curvature components are non-vanishing
since the Christoffel symbols have components of the
form $\Gamma^i_{\jbar k} = \frac{1}{2} G^{i \bar{\ell}} \left( \pd_{\jbar} G_{\bar{\ell} k} - \pd_{\bar{\ell}} G_{\jbar k} \right)\neq0$.}

However, we
will see in the next section that all of the contributions to the moduli space metric coming from the curvature are in fact vanishing to $O(\ap)$.  Thus, we find that (somewhat surprisingly) the K\"ahler and complex structure moduli remain decoupled through $O(\ap^2)$. It is tempting
to speculate whether or not this decoupling persists to higher orders in $\ap$. At first sight this might seem unlikely since the vanishing of~\C{mixed metric}\
does not seem to extend beyond $O(\ap^2)$. However, higher order corrections may conspire to maintain a direct product moduli space. While
such a proposal may seem far-fetched, it is not entirely unfounded in light of $(0,2)$ mirror symmetry.


\subsection{The K\"{a}hler potential for K\"{a}hler moduli}

Let us now examine the metric for the K\"{a}hler moduli,
\be
\cG_{\alpha \bar{\beta}} = {1\over 2\cV\,'} \int d^6y \sqrt{G} \dd_{\a} G_{i\bar{\ell}}\, \dd_{\bar{\beta}} G_{\jbar k}\, \Big(G^{i\jbar}G^{k\bar{\ell}}-{\ap\over2}\cR^{i\jbar k\bar{\ell}}\Big) \label{kahler metric},
\ee
with the aim of understanding the associated space-time K\"{a}hler potential. We begin by considering the ${O} (\alpha')$ deformation, expanding $G_{i \jbar} = g_{i \jbar} + \alpha' h_{i \jbar}$. As discussed in section~\ref{(2,2)review}, the K\"{a}hler form of a Calabi-Yau space can be expressed as a linear combination of integral 2-forms:
$J = t^\a\o_\a$. The $t^\a$ are the spacetime moduli fields, while the
\be
\o_\a = i\left(\dd_\a g_{i\jbar}\right)\d y^i\wedge\d y^\jbar
\ee
form a basis of $H^2(\M,\Z)$. Because $h^{2,0}=0$, these forms span $H^{1,1}(\M)$. Similarly, from $h_{i\jbar}$ we can construct a two-form
\be
\tilde{h} = i\, h_{i\jbar}\, \d y^i\wedge \d y^\jbar \, .
\ee
Following the discussion in section~\ref{background-solution}, the form $\tilde{h}$ is orthogonal to all harmonic 2-forms: if $\g\in\bigwedge^2T^*\cM$ and
$\D\g=0$, then
\be
\int \tilde{h}\wedge*\g = 0  .
\ee
Varying the moduli changes both $g \rightarrow g + \delta g$ and $h \rightarrow h + \delta h$. We will argue now that the associated two-form $\delta \tilde{h}$ is also orthogonal to harmonic forms, just like $\tilde{h}$. First, recall that the equation of motion for $h_{mn}$ is
\be
\D_L h_{mn}= \hlf\left[ \tr(F_{mp}F_n^{\ p}) - R_{mpqr}R_{n}^{\ pqr} \right] \, , \label{heom3}
\ee
where $F_{mp}$ and $R_{mnpq}$ are the zeroth order quantities. After deforming the solution
$$g + \alpha' h \,\rightarrow \,(g+\dd g) + \alpha' (h + \dd h)$$
the right hand side of~(\ref{heom3})\ changes along with the operator $\D_L\simeq-\nab^2-R$ that depends implicitly
on $g$. Hence $h_{mn}+\dd h_{mn}$ is a solution of the deformed \C{heom3}, provided
\be
\D_L \dd h_{mn} = -(\dd \D_L) h_{mn} + \dd \left( \hlf\tr(F_{mp}F_{nq})g^{pq} - \hlf R_{mpqr}R_n^{\ pqr} \right) \neq 0 \, .
\ee
The exact functional form on the right-hand-side of this equation is not important. The essential point is that it is non-vanishing and, thus, $\dd h_{mn}$ must be
orthogonal to the zero-modes of Lichnerowicz just like $h_{mn}$.

Let us now rewrite the moduli space metric in terms of the above forms. With a little work, we find that to order $\alpha'$ the corrected metric takes the form
\be \label{Gab}
\cG_{\alpha \bar{\beta}} = {1\over 2\cV} \int\Big[\o_\a\wedge*\o_{\bar{\beta}} + \ap \Big(\dd_\a\tilde{h}\wedge*\o_{\bar{\beta}} + \o_\a\wedge*\dd_{\bar{\beta}}\tilde{h} -\o_\a\wedge\o_{\bar{\beta}}\wedge\tilde{h}-
{\sqrt{g}\over2}R(\o_\a,\o_{\bar{\beta}}) \Big)\Big]
\ee
where $R(\o_\a,\o_{\bar{\beta}})$ denotes $R^{i\jbar k\bar{\ell}}\o_{\a \,i\bar{\ell}}\,\o_{\bar{\beta} \,\jbar k}$\,.

Because of N=1 space-time supersymmetry, we know that the moduli space metric must be K\"{a}hler and so it should be possible to derive it from a K\"{a}hler potential. A straightforward, but tedious, exercise shows that
\be
\cG_{\alpha \bar{\beta}} = -\hlf{\del\over\del t^\a}{\del\over\del t^{\beta}}\log \int\left( J\wedge J\wedge J +3\ap J\wedge J\wedge \tilde{h} + \ap\sqrt{g}R\right) +O(\ap^2) \,\, .
\ee
Since the scalar curvature in the last term of the bracket should be computed with the lowest order metric $g$, we see that to this order $\alpha' \sqrt{g} R$ vanishes identically.
This is far from obvious if we only look at the term appearing in the metric $R(\o_\a,\o_{\bar{\beta}})$. However, once this metric is expressed in terms of a K\"{a}hler potential, it becomes clear. It would be interesting to understand whether one could  prove directly that
$\int R(\o_\a,\o_{\bar{\beta}})\equiv0$, perhaps by using some properties of CY topological invariants.

So far, we have found that the $\ap$-corrected K\"{a}hler potential is:
\be
K = -\log \int\left( J\wedge J\wedge J +3\ap J\wedge J\wedge \tilde{h} \right) +O(\ap^2) = -\log\int J'\wedge J'\wedge J' \label{kahlerpot} \,\, ,
\ee
where $J'=J+\ap\tilde{h}+\,\ldots$ is the corrected fundamental form. At this point, it is obvious that the whole ${O} (\alpha')$ correction vanishes because of~(\ref{htorth}), that is to say the orthogonality of $\tilde{h}$ and $J$.

However, with a view to generalizing this result to higher orders in $\alpha'$, it is useful to make a few observations about the form of~(\ref{kahlerpot}). Note that the K\"ahler potential can be written in the form $K=-\log \cV\,'$, as was the case for the classical K\"ahler potential.
Actually, given the form of~\C{kahler metric}, it is clear in hindsight that the corrected moduli space metric should simply be obtained by replacing
$J\rightarrow J'$ in the expression for the uncorrected K\"{a}hler potential~(\ref{(2,2)potential}).


Before we turn to $O(\alpha'^2)$, let us pause for a moment and note that the vanishing of the ${O} (\alpha')$ correction is unexpected since there is no obvious space-time reason to expect such a vanishing. Indeed, since there are $O(\alpha')$ corrections in the ten-dimensional action, we might have expected ${O} (\alpha')$ corrections in the 4-d effective action as well.
It would certainly be very interesting to understand whether there is a deeper reason for this vanishing or whether this is just accidental.


Having shown that the ${O} (\ap)$ correction vanishes, we will now determine the leading behaviour of the K\"{a}hler potential. Based on the
form of the metric \C{kahler metric}, it is natural to expect that the structure of \C{kahlerpot} will persist to higher orders. Indeed, writing
\be \label{Jprime}
J' = J +\ap \tilde{h} +\ap^2 \tilde{h}\2 + \,\ldots \,,
\ee
where $\tilde{h}\2$ is the $(1,1)$ form associated to the metric correction $h\2_{i\jbar}$ in (\ref{GmnExp}), one can show that the K\"{a}hler potential to $O(\ap^2)$ is given by
\be \label{Kf2}
K = -\log\int \left(J'\wedge J'\wedge J' +\ap\sqrt{G}\cR\right).
\ee
See Appendix~\ref{curvature}\ for the details of this computation.
The last term is still vanishing for the following reason. Expanding out the fields gives
\be \label{GcalR}
\ap\sqrt{G}{\cal \cR} = (\ap/2)\sqrt{g}\left(2R + \ap h R -\ap\nab^2h\right) + O(\ap^3).
\ee
Since $R=0$, we are left with a total derivative which vanishes on integration over the compact space. On the other hand, expanding the first term in~(\ref{Kf2}), we find
\be
 J\wedge J\wedge J +3\ap J\wedge J\wedge \tilde{h} + 3\ap^2 J\wedge J \wedge\tilde{h}\2 + 3\ap^2 J\wedge \tilde{h}\wedge\tilde{h} +O(\ap^3) \,\, .
\ee
The terms linear in $\tilde{h}$ and $\tilde{h}\2$ vanish by orthogonality to harmonic forms leaving
\be
K = - \log (\cV ) - {\ap^2\over 2\cV}\int J \wedge \tilde{h} \wedge \tilde{h} +O(\ap^3) = - \log (\cV ) +{\ap^2\over 2\cV}\int \tilde{h}\wedge*\tilde{h} +O(\ap^3). \label{kahlerpot-final}
\ee
We have used $*\tilde{h} = - J \wedge \tilde{h} + \frac{3}{2} \frac{\int \tilde{h} \wedge J \wedge J}{\int J \wedge J \wedge J} \,J \wedge J$ in~\C{kahlerpot-final}.

To summarize: the leading correction to the K\"{a}hler potential appears at $O(\ap^2)$ and is controlled by the norm-squared of $\tilde{h}$. In particular, this
correction vanishes if and only if $\tilde{h}$ vanishes, which occurs in the case of the standard embedding.\footnote{It is reasonable to expect that smooth bundle deformations away from the standard embedding solution will also preserve this vanishing.}

To conclude this section, let us check the volume dependence of the terms in~\C{GmnExp}, or equivalently~\C{Jprime}. The leading order term scales like ${\cal V}^{1/3}$ by definition. The expansion is in powers of ${\ap\over {\cal V}^{1/3}}$ so $h$ is independent of ${\cal V}$ while $h^{(2)}$ is suppressed by ${\cal V}^{-1/3}$. Defining ${\cal V} = e^{6u}$, we can make manifest the $u$-dependence of the background metric\footnote{The notation we use here is for easy comparison with the reduction ansatz of~\cite{Anguelova:2005jr}.}
\be\label{metricscaling}
ds^2 = \hat{g}_{\mu \nu} dx^{\mu} dx^{\nu} + \left( e^{2u(x)} \tilde{g}_{mn} + \alpha' h_{mn} + \alpha'^2 e^{-2u(x)} h^{(2)}_{mn} + \ldots \right) dy^m dy^n.
\ee
That $h_{mn}$ is independent of $u$ will be important in section~\ref{Implications}.



\section{Breaking No-Scale Structure}\label{Implications}
\setcounter{equation}{0}

In this section, we will examine the implications of the $\ap$-corrected K\"ahler potential \C{kahlerpot-final} in proposed moduli stabilization scenarios.
As we explained in the introduction, most discussions of moduli stabilization occur in type IIB or F-theory compactifications where the difficult modulus
to stabilize is the volume modulus. The remaining moduli can all  be stabilized, in principle, by fluxes~\cite{Dasgupta:1999ss}. To find reliable fully
stabilized vacua of either supersymmetric or non-supersymmetric type will, generically, require knowledge of string scale physics.

Nevertheless, there have been various proposed scenarios based on four-dimensional effective field theory. For reasons mentioned in the introduction, we
will take an agnostic view about the relation of those scenarios to string theory and simply ask how the effective field theory analysis changes taking
into account our results.

Assuming some superpotential $W$, we can ask whether the $(\alpha')^2$ correction to $K$ gives the leading breaking of the no-scale structure.  Recall that
the supergravity scalar potential is given in terms of $K$ and $W$:
\be
V = e^{K}\left[K^{{\cal I}\bar{{\cal J}}}D_{\cal I} W D_{\bar{{\cal J}}} \bar{W} - 3|W|^2\right].\label{V}
\ee
We are not concerned about the explicit moduli-dependence of $W$ since that depends on detailed scenarios.
What really interests us is whether the combination,
\be \label{noscale}
V   = e^K\left(
K^{{\cal I}\bar{{\cal J}}} K_{{\cal I}} K_{{\bar{{\cal J}}}} - 3 \right)|W|^2 + \ldots,
\ee
is zero or non-zero for our correction term.\footnote{Recall that this combination vanishes for the classical K\"ahler potential
$K=-\log(\cV)$. This results in no scalar potential for the overall volume modulus; hence the term ``no-scale structure."}

First let us examine the heterotic string and imagine a superpotential $W$ generated by non-perturbative physics and perhaps by background fluxes. Flux induced
superpotentials in the heterotic string have been described largely using duality with type IIB backgrounds in~\cite{Becker:2003yv, Becker:2003gq, LopesCardoso:2003af, Held:2010az}.
These superpotential discussions are on a somewhat less secure footing than the type IIB case because the full space of chiral fields on which the superpotential
and K\"ahler potential depend are not visible in supergravity.

Let us take a simple one K\"ahler modulus case with classical $K = - 3 \log (T - \bar{T} )$. In this case, we
can factor the $T$-dependence out of the quantum correction to give
\be
K = - 3 \log (T - \bar{T} ) + {\ap^2 \over (T-\bar{T})^{2} } \times f
\ee
with $f$ independent of $T$. Computing the leading terms in the potential gives
\be\label{hetpotential}
V =   - {2 \ap^2 f \over t^2 }  e^K |W|^2 + \ldots.
\ee
What we want to note is that this is non-vanishing at leading order in the correction. Models of this type will be discussed in greater detail in a future publication~\cite{Anguelova:2010qd}.

Now we want to consider the implications for type I and type IIB. The first step is to map the heterotic correction to type I using ten-dimensional
S-duality. By definition, the Einstein frame physics is invariant under S-duality so what we care about is the form of the correction in type I string
frame. Using standard relations (see, for example~\cite{Antoniadis:1997nz}) gives a correction to $K$ of the form,
\be\label{typeIcorr}
({\tilde g_s})^2 { \ap^2 \over {\tilde \cV}^{2/3}} \times {\tilde f},
\ee
expressed in terms of the type I string coupling, ${\tilde g_s}$, and volume ${\tilde \cV}$. Again the moduli-dependence is shuffled into a function
${\tilde f}$. This is loop suppressed in type I string frame.

 It is natural that this should be the case because the type I origin of the $\ap R^2$ terms
in ten dimensions is open strings whose interactions are suppressed by the string coupling. Given that the classical K\"ahler potentials for $T$ and $S$ are
the same in type I and heterotic, it follows that the leading no-scale structure breaking will be functionally similar. In particular, it will begin at $O(\ap^2)$.

Now we might wonder if F-theory type orientifolds, as opposed to type I, might have different leading corrections to $K$. It is easy to see that this
is not the case both on general grounds and by considering special cases like quotients of $K3\times T^2$. In the latter models, T-duality along the $T^2$ maps
us between F-theory orientifolds and type I but the quantum corrections are independent of the volume of $T^2$ thanks to the perturbatively exact isometries. We again
expect $O(\ap^2)$ corrections with string loop suppression.  This is also in agreement with the explicit computations of~\cite{Berg:2005ja, Berg:2007wt}.

Surprisingly, this does not mean that the breaking of no-scale structure takes the same form in F-theory/IIB orientifolds as in heterotic/type I. In heterotic and type I,
the natural variables in which to express the low-energy theory are $(T^\a,S)$. In F-theory type orientifolds, the natural variables are $(\rho_\a, \tau)$
where $\rho_\a$ measure the volumes of $4$-cycles and $\tau$ is the complexified type IIB coupling. We would like to map the corrections we derived in heterotic
over to F-theory orientifolds using a duality map. This is not a trivial exercise because only chiral field redefinitions of chiral fields are permitted in the space-time effective action so the map of variables from one description to a dual description is quite subtle.
See~\cite{DAuria:2004kx}\ for a discussion of the issues one encounters. The upshot is that the map itself receives quantum corrections which is an added
complication. Nevertheless, there is a very reasonable conjecture for the resulting K\"ahler potential based on our preceding discussion:
\be\label{iibkahler}
K_{IIB} = -2\log\left[\cV\right] + \ap^2 \left({2i\over  {\t-\bar{\t} }}\right) {f_{IIB}\big(\rho_\a-\bar{\rho}_\a\big)\over\cV^{2/3}}   - \log[\t-\bar{\t}] .
\ee
The $g_s$ dependence differs from the (string frame) type I result \C{typeIcorr} only as a result of transforming to Einstein frame. Otherwise, it is
functionally the same. At this stage, the function $f_{IIB}$
is some unknown function that depends on all K\"ahler moduli \textit{except} the overall volume.\footnote{In the special cases of $K3\times T^2$ quotients,
we would find $f_{IIB} = 3\cV^{-1/3}\int J\wedge\tilde{h}\wedge\tilde{h}$. However for more general cases, the form of $f_{IIB}$ could be more complicated.}
Another way to say this is that $f_{IIB}$ is a function of the scale-invariant variables\footnote{We wish to thank M. Cicoli and J. Conlon
for stressing this point.}
\be
\hat{\rho}_\a = {\rho_\a\over\cV^{2/3}}.
\ee
These new variables are homogeneous functions of degree 0 in the original $\rho$ variables, thereby ensuring that $f_{IIB}$ is independent of the overall scale, as required. Many authors
have studied corrections to the IIB K\"ahler potential of this form (with $f_{IIB}$ being degree 0 in $\rho_\a$)~ \cite{Berg:2005ja,Berg:2007wt,Cicoli:2007xp}, and they all find that the $O(\ap^2)$ correction to
the scalar potential vanishes! In fact, it is straightforward to compute the leading no-scale breaking term using \C{iibkahler}. This yields
\be
V = - {g_s \ap^2 \over \cV^{8/3}}|W|^2 \rho_\a \bar{\rho}_\beta {\del^2 \over \del{\rho_{\a}}\del{\bar{\rho}_{\beta}}} f_{IIB}\big(\hat{\rho}_\a - \hat{\bar{\rho}}_\a\big)  +\ldots
\ee
which is easily seen to vanish because $f_{IIB}$ is of degree 0 with respect to the $\rho_\a$ variables.
This unexpected cancelation was dubbed ``extended no-scale structure" in \cite{Cicoli:2007xp}. Therefore the leading $\ap$ correction to the scalar potential in type IIB
does not appear until $O(\ap^3)$. This is in sharp contrast to the heterotic/type I result that we described above.

\subsection*{Acknowledgements}

It is our pleasure to thank K. Becker, M. Cicoli, J. Conlon, M. Haack, J. Louis, A. Parnachev and N. Saulina for useful conversations. We would also like to thank the organizers and participants of the BIRS workshop on ``(0,2) Mirror Symmetry and Heterotic Gromov-Witten Invariants" for providing a stimulating atmosphere. S.~S. would also like to thank the University of Amsterdam for hospitality during the completion of this project.

L.~A. is supported by DOE grant FG02-84-ER40153. C.~Q. is supported in part by NSF Grant No.~PHY-0758029 and by an NSERC PGS-D research scholarship. S.~S. is supported in part by
NSF Grant No.~PHY-0758029, NSF Grant No.~0529954 and the Van der
Waals Foundation.

\appendix

\section{Reduction of the Action} \label{reduction}
\setcounter{equation}{0}
Here we collect the details of the reduction of the ten-dimensional action (\ref{HetAc}) to four dimensions. Using the ansatz from Section \ref{step1},
we will treat the fields exactly, and postpone expanding in $\ap$ until the end. To avoid possible confusion we will label ten-dimensional quantities
with a ${(10)}$ subscript.

Let us begin by decomposing the Ricci scalar. With our ansatz for the metric, the non-vanishing components of the Levi-Civita connection are:
\bea
\G_{(10)}{}_{\m\n}^{\la} &=& \widehat{\G}_{\m\n}^\la, \non\\
\G_{(10)}{}_{\m m}^n &=& \hlf G^{np}\del_\m G_{mp}, \label{connection}\\
\G_{(10)}{}_{mn}^\m &=& -\hlf \del^\m G_{mn}, \non\\
\G_{(10)}{}_{mn}^p &=& \G_{mn}^p .\non
\eea
Here the $\hat{\G}$ components are built from the four-dimensional metric $\hat{g}$.
Using the above connection, it is straightforward to work out the Ricci scalar
\be
R_{(10)} = \widehat{R} + G^{mn}R_{mn} - \widehat{\nab}^2 \log|G| -{1\over4}\left(\del_\m \log|G|\right)^2 -{1\over 4}G^{mn}G^{pq}\,\del_\m G_{mp}\,\del^\m G_{nq},
\ee
where hatted quantities are constructed from the space-time metric $\hat{g}$.

Next, let us consider the dilaton kinetic term. It has the simple form:
\be
\big(\del_M \Phi_{(10)}\big)^2 = \big(\del_\m \Phi_{(10)}\big)^2 + \big(\del_m \f\big)^2 \, .
\ee
Rather than split the fluctuation $\varphi$ from the background $\f$, as in \eq{dilatonsplit}, it will prove
convenient to keep these combined in the full field $\Phi$.

We move now to the $\H$ field. Noting that it includes the Chern-Simons couplings, we find:
\be
\big| \H_{(10)}\big|^2 = {1\over6}\H_{\m\n\la}\H^{\m\n\la}  + {1\over6}\H_{mnp}\H^{mnp} \, ,
\ee
where
\bea
\H_{\m\n\la} &=& 6\Big(\del_{[\m} B_{\n\la]} -{\ap\over4}\big(A_{[\m} F_{\n\la]} - {2\over3}A_{[\m} A_\n A_{\la]} \big)\Big) + O(\del^3) \label{H-components} \, .
\eea
Components with mixed space-time and internal indices do not appear since we are neglecting the $B$-field and bundle moduli as well as the
charged matter fields in this analysis. This simplification carries over to the gauge sector as well. The gauge kinetic terms are simply
\bea
\Tr\big|\cF_{(10)}\big|^2 &=&  \Tr\Big(\, \hlf F_{\m\n}F^{\m \n} + \hlf \cF_{mn}\cF^{mn} \Big) \, , \non
\eea
where Tr is taken in the adjoint representation of $\cG = E_8\times E_8$ or $SO(32)$.

Finally, we consider the curvature-squared terms. Since we are only dealing with bosonic fields, the spin connection is not really needed
so we simply use the (torsionful) Levi-Civita connection, suitably modified from~\eq{connection}.
Most terms in $\cR_+^2$ lead to higher derivative space-time couplings, but since we are only concerned with the two derivative action,
these may be neglected safely. The only
relevant component of the curvature tensor is\footnote{We neglect the component $\cR_{+\,mnp}^{\m}$, although it leads to a two-derivative term in the action, for the following reason. One can easily show that:
\be
\cR_{+\,mnp}^{\m}= -2\nab_{[m}\G_{n]p}^{\m} = -\del^{\m}M_{\cal I}(x)\nab_{[m}\dd^{\cal I} G_{n]p} = O(\ap) \, \nn .
\ee
Therefore, $\ap \cR_{+\, mnp\m}^2\sim O(\ap^3)$, which is beyond the order of interest in this paper.}
\bea
\cR_{+_{(10)}\, mnpq} &=& \cR_{+\, mnpq} - \hlf \del_\m G_{p[m}\del^\m G_{n]q} \, ,
\eea
where $\cR_{+\, mnpq} = \cR_{mnpq} + (d\H)_{mnpq}$. This contributes to the action via
\bea
\tr\big|\cR_+\big|^2 &=&  \hlf \cR_{+_{(10)}\,mnpq}\cR_{+_{(10)}}^{mnpq} + O(\del^4), \\
&=& \hlf \cR_{+\, mnpq}\cR_+^{mnpq} - \hlf\del_\m G_{mp}\del^\m G_{nq}  \cR^{mnpq} .\non
\eea
Notice that $d\H$ drops out of the last term because of its index structure.

Collecting this results, we obtain the following ten-dimensional action for the decomposed fields:
\bea
S &=& {1\over 2\k_{10}^2} \int\d^{10}x\sqrt{-\hat{g}}\sqrt{G}e^{-2\Phi}\Big[\hat{R}-\widehat{\nab}^2\log|G|-\qrt\big(\del_\m \log|G|\big)^2 +
4(\del_\m \Phi)^2 -{1\over12}H_{\m\n\la}^2  \non\\
&&~~~~~~~~~~~~~~ - {\ap\over8}F_{\m\n}^2-\qrt\del_\m G_{mp}\del^\m G_{nq}\Big(G^{mn}G^{pq} +{\ap\over2}\cR^{mnpq}\Big) +O(\ap^3) \Big] \, . \label{action1}
\eea
Note that in writing this action, we have thrown away the purely internal contribution
\be
\int\d^6y\sqrt{G} e^{-2\Phi}\Big[G^{mn}\cR_{mn} + 4(\del_m\f)^2 -{1\over12}\H_{mnp}\H^{mnp} -{\ap\over8}\big(\Tr\cF_{mn}\cF^{mn} - \cR_{+\, mnpq}\cR_+^{mnpq}\big) \big] , \non
\ee
which would seem to lead to a potential for the moduli fields.
However, it actually vanishes because of the equations of motion; specifically, because of the dilaton field equation in~(\ref{eom}).

A simple way to see this is as follows:
consider the full effective Lagrangian coming from tree-level string theory with its infinite set of higher derivative corrections. The key is that,
in string frame, the dilaton-dependence is homogeneous. For clarity, let us write
\be
\L_{tree}(\Phi, \del\Phi,\ldots)=e^{-2\Phi}\tilde{\L}_{tree}(\del \Phi,\ldots),
\ee
where we have separated the (non-derivative) dilaton-dependence from the rest of the action. The ellipsis denote the rest of the fields appearing in the action.
The dilaton equation of motion then becomes
\be
\L = -\hlf \del_N{\dd \L \over \dd \del_N\Phi} .
\ee
The string-frame action therefore always vanishes when evaluated on a static classical solution. This is why we can neglect terms involving only
the background fields $g_{mn}$, $B_{mn}$, $A_m$, $\f$ and their derivatives with respect to \textit{internal} directions. On the other hand,
\textit{external} derivatives of these fields do, through their implicit $x$-dependence, lead to kinetic terms for moduli and so cannot be omitted.

Although, in principle, we neglect the matter fields $C$, it is interesting to see how this argument would be modified in their presence.
Since the vacuum solution corresponds to $C=0$, nothing precludes the possibility of a potential that vanishes in this limit.
Indeed, $|F|^2$ contains the non-derivative terms $[A,A]^2$, which do generate a potential for the $C$'s. This leads to the well-known superpotential
\be
W = \e^{ijk}\Tr C_iC_jC_k ,
\ee
where the $C_m$ fields are in a complex basis and viewed as N=1 chiral superfields.

Now let us return to~\eq{action1}. In order to obtain a four-dimensional effective action, we first define the four-dimensional dilaton by,
\be \label{DilDef}
e^{-2\f_4(x)} = {1\over \cV_0}\int\d^6y\sqrt{G}e^{-2\Phi} ,
\ee
where $\cV_0$ is some reference volume.\footnote{As usual, $\cV_0$ gets absorbed into the four-dimensional Newton constant (which we set to 1) along with the dilaton zero-mode:  $\k^{-2}= \cV_0e^{-2\f_{_0}}\k_{10}^{-2}=1$.}
Note that this is the standard definition in both classically non-K\"{a}hler and more general $SU(3)\times SU(3)$ structure compactifications; see, for instance,~\cite{Cassani:2007pq}.
In our case, since the non-K\"{a}hlerity is due to $\alpha'$ effects, definition~(\ref{DilDef}) contains an infinite number of $\ap$ corrections compared to the classical one given by
$$\f_4(x) = \varphi(x) -\hlf \log \left[ { \cV(x)\over \cV_0} \right]$$
with $\cV(x)= \int\d^6y\sqrt{g}$ the CY volume. This definition is also consistent with~\cite{Becker:2002nn}\ where it was shown that the $\alpha'^3 R^4$ term in the ten-dimensional effective action also leads to an $\alpha'$-dependent correction to the definition of the 4-d dilaton.
Using~(\ref{DilDef}), we will
obtain the proper gravitational action in Einstein frame, after performing the standard Weyl transformation
\be
\hat{g}_{\m\n}=\exp(2\f_4)g_{E\, \m\n}
\ee
of the four-dimensional metric.
Finally, we have to dualize $B_{\m\n}$ in favour of an axion scalar field $a$. Because
of the modified Bianchi identity, we must add the Lagrange multiplier
\be
{1\over2} \int  a\left[\d\H + {\ap\over4}\Tr F\wedge F \right]
\ee
to the action and integrate out $H_{\m\n\la}$. As usual, we combine this axion with $\f_4$ into a complex scalar
\be
S = a + i e^{-2\f_4} \, .
\ee

After performing all of the above steps, we arrive at the following four-dimensional effective action:
\bea
S &=& {1\over2}\int\d^4x\sqrt{-g_E}\left[R_E+2{\del_\m S \del^\m\bar{S} \over (S-\bar{S})^2} -\cG_{{\cal IJ}}(M)\del_\m M^{\cal I} \del^\m M^{\cal J} \right]\label{eff-action2}\\
&& -{\ap\over8}\int \Tr\Big(\Im\!(S)F\wedge*F -\Re\!(S)F\wedge F\Big) +O(\ap^3)  , \non
\eea
where the moduli space metric is\footnote{We should mention two simplifications used here, which only hold in our preferred $\xi=0$ and $\zeta=1$ gauge.
First, the term
\be
2\left({e^{2\f_4}\over \cV_0}\right)^2\int\d^6y\d^6y'\sqrt{G(y)}\sqrt{G(y')}e^{-2\Phi(y)}e^{-2\Phi(y')}\dd_{\cal I}\big(\Phi(y)-\Phi(y')\big)\dd_{\cal J}\big(\Phi(y) -\Phi(y')\big) \non
\ee
has been dropped because in this gauge, the $y$-dependence of $\Phi$ begins at $O(\ap^3)$ and so this term is in fact $O(\ap^6)$. Secondly, the volume pre-factor $1/\cV'$
is the remnant of the dilaton-dependent term:
\be
{e^{2\f_4-2\Phi}\over\cV_0} = \Big(\int\d^6y'\sqrt{G}e^{-2\Phi(y')+2\Phi(y)}\Big)^{-1} = {1\over\cV'}+O(\ap^3) .\non
\ee}
\bea
\cG_{{\cal IJ}} &=& {1\over 4\cV\,'} \int d^6y \sqrt{G}\,\dd_{\cal I} G_{mp}\, \dd_{\cal J} G_{nq}\, \Big(G^{mn}G^{pq}-{\ap\over2}\cR^{mnpq}\Big)  . \label{metric1}
\eea
Recall that the variations $\dd_I = {\dd\over\dd M^I}$ are with respect to the moduli fields. The $\ap$-corrected volume $\cV'$ is defined by,
\be
\cV' = \int\d^6y\sqrt{G} = \int J'\wedge J' \wedge J' \, ,
\ee
where $J'_{i\jbar} = iG_{i\jbar} = i(g_{i\jbar} + \ap h_{i\jbar} + \ldots)$ is the corrected fundamental form. Finally, we point out that by using the
generalized $\xi=0$, $\zeta=1$ gauge that eliminates $\f\2$, the 4-d dilaton acquires the form:
\be
\f_4 = \varphi - \hlf \log \left[ {\cV'\over \cV_0}\right] + O(\ap^3),
\ee
which still differs from the classical expression by the replacement of $\cV\mapsto\cV'$.

\section{Curvature Term in the K\"ahler Potential} \label{curvature}
\setcounter{equation}{0}

Showing that the metric \C{metric3} descends from the K\"{a}hler potential \C{Kf2} involves some subtle calculations, especially regarding the curvature term.
The main difficulty lies in reducing
\bea
\dd_{\cal I}\dd_{\cal J}\int\ap\sqrt{G}\cR &=& -{\ap\over2}\int\sqrt{G}\Big(\cR^{mnpq}-\hlf G^{mn}G^{pq}\cR + G^{mp}G^{nq}\cR + \\
&&\qquad\qquad~~~~~ + G^{mp}\cR^{nq} + G^{mn}\cR^{pq} + G^{pq}\cR^{mn}  \Big)\dd_{\cal I} G_{mn} \dd_{\cal J} G_{pq} \non\\
&& + \ap\int\sqrt{G}\big(G^{mn}G^{pq}-G^{mp}G^{nq}\big)\big(\dd_{\cal I} G_{mn}\D_L \dd_{\cal J} G_{pq} +\dd_{\cal J} G_{pq}\D_L\dd_{\cal I} G_{mn} \big) \non
\eea
to only the first term in the first line:
$$-{\ap\over2}\int\sqrt{G}\cR^{mnpq}\dd_{\cal I}G_{mp}\dd_{\cal J}G_{nq}.$$
The extra
terms can be expanded in $\ap$ yielding:
\bea \label{extra}
\hspace*{-0.65cm}&&-{\ap^2\over2}\!\int\!\!\sqrt{g}\Big(\qrt g^{mn}g^{pq}\nab^2 h -\!\hlf g^{mp}g^{nq}\nab^2 h + g^{mp}\D_L h^{nq} \!+ g^{mn}\D_L h^{pq} \!+ g^{pq}\D_L h^{mn} \!\Big)\dd_{\cal I} g_{mn} \dd_{\cal J} g_{pq} \non\\
\hspace*{-0.65cm}&& + \ap^2\!\int\!\!\sqrt{g}\big(g^{mn}g^{pq}-g^{mp}g^{nq}\big)\!\big(\dd_{\cal I} g_{mn}\D_L \dd_{\cal J} h_{pq} +\dd_{\cal J} g_{pq}\D_L\dd_{\cal I} h_{mn} \big) +O(\ap^3) .
\eea
Notice that $g^{mn}\dd g_{mn}$ is actually a constant (which follows directly from taking trace of $\D_L \dd g_{mn}=0$). Because of this and the fact that
$h_{mn}$ and $\dd h_{mn}$ are orthogonal to $\dd g_{mn}$, it is easy to see that most terms in (\ref{extra}) vanish. The only ones that are left at this point are the second and
third terms in the first line. They can be written as
\bea \label{lastextra}
\int\sqrt{g}\, \D_L \bar{h}^{nq} \,\dd_{\cal I} g_{mn} g^{mp} \dd_{\cal J} g_{pq} = (\D_L \bar{h}\,, \dd_{\cal I} g\cdot\dd_{\cal J} g)  ,
\eea
where
\be
\bar{h}_{mn} = h_{mn} + g_{mn} h
\ee
and $(~\,,~)$ is the obvious inner-product for symmetric 2-tensors.

To show the vanishing of~(\ref{lastextra}), we note that the Lichnerowicz operator can be defined
via $\D_L = \nab_S^*\nab_S$, where $\nab_S$ is the symmetrized covariant derivative
$$\nab_{S}h = \nab_{(m}h_{np)},$$
and $\nab_S^*$ is its formal adjoint:
$(\nab_S^*h_1,h_2)=(h_1,\nab_S h_2)$~\cite{Gibbons:1978ji}. Note also that zero-modes of Lichnerowicz are  zero-modes of $\nab_S$ as well. This is easy to see by considering
a zero-mode $\dd g$ of Lichnerowicz. Then one notes
$$0=(\dd g,\D_L\dd g) = (\nab_S \dd g,\nab_S\dd g).$$
Combining the above facts, we find:
\be
 (\D_L \bar{h}\,, \dd_{\cal I} g\cdot\dd_{\cal J} g) = \big(\nab_S \bar{h}\,, \nab_S(\dd_{\cal I} g\cdot\dd_{\cal J} g)\big) =0,
\ee
where we have used the Leibniz rule to see that $\dd_{\cal I} g\cdot\dd_{\cal J} g$ is also a zero-mode of $\nab_S$. This completes the proof.


\newpage


\ifx\undefined\bysame
\newcommand{\bysame}{\leavevmode\hbox to3em{\hrulefill}\,}
\fi

\end{document}